\documentclass[english,prl,twocolumn,noeprint,longbibliography,superscriptaddress]{revtex4-1}
\usepackage[T1]{fontenc}
\usepackage[latin9]{inputenc}
\usepackage{color}
\usepackage{babel}
\usepackage{amsmath}
\usepackage{amssymb}
\usepackage{graphicx}
\def\ket#1{| #1 \rangle}

\usepackage[unicode=true,pdfusetitle,
 bookmarks=true,bookmarksnumbered=false,bookmarksopen=false,
 breaklinks=true,pdfborder={0 0 0},pdfborderstyle={},backref=false,colorlinks=true]
 {hyperref}

\begin{document}

\title{Neural Belief-Propagation Decoders for Quantum Error-Correcting
  Codes}

\author{Ye-Hua Liu}

\affiliation{D\'epartement de Physique \& Institut Quantique,
  Universit\'e de Sherbrooke, J1K 2R1 Sherbrooke, Qu\'ebec, Canada}

\author{David Poulin}

\affiliation{D\'epartement de Physique \& Institut Quantique,
  Universit\'e de Sherbrooke, J1K 2R1 Sherbrooke, Qu\'ebec, Canada}

\affiliation{Canadian Institute for Advanced Research, M5G 1Z8
  Toronto, Ontario, Canada}

\begin{abstract}
  Belief-propagation (BP) decoders play a vital role in modern coding
  theory, but they are not suitable to decode quantum error-correcting
  codes because of a unique quantum feature called error
  degeneracy. Inspired by an exact mapping between BP and deep neural
  networks, we train neural BP decoders for quantum low-density
  parity-check (LDPC) codes with a loss function tailored to error
  degeneracy. Training substantially improves the performance of BP
  decoders for all families of codes we tested and may solve the
  degeneracy problem which plagues the decoding of quantum LDPC codes.
\end{abstract}

\maketitle

Statistical inference on a graph is an important paradigm in many
areas of science, and equivalent heuristic algorithms have been
developed by different communities, including the cavity method in
statistical physics \citep{Mezard1986} and the belief propagation (BP)
algorithm in information science \citep{Pearl1982}. In the latter
case, BP is the standard decoding algorithm for low-density
parity-check (LDPC) codes \citep{Gallager1962}, which form the
backbone of modern coding theory and are widely used in wireless
communication \citep{Richardson2008}.  With the growing interest for
quantum technologies, quantum generalizations of LDPC codes have been
proposed \citep{MacKay2004,Gilles2009,Leverrier2015}, but BP was found
to be inadequate for their decoding \citep{Poulin2008} because of
error degeneracy, a feature unique to quantum codes.  Despite many
improvements \citep{Poulin2008,Wang2012a,Babar2015} to BP, there is
still no accurate decoding algorithm for general quantum LDPC
codes. This contrast with statistical physics where the cavity method
has been generalized to the quantum setting with some success
\citep{Hastings2007,Poulin2008a,Laumann2008,Poulin2011b}.

Recently, an exact mapping between BP and artificial neural networks
has been revealed \citep{Nachmani2016}, which implies a general
machine-learning strategy to adapt BP to any specific task. In this
article, we use this strategy for the decoding of quantum LDPC
codes. Neural-network-based decoders for quantum error-correcting
codes have attracted great interest recently, particularly in the
context of topological codes
\citep{Torlai2017a,Baireuther2017,Krastanov2017,
  Varsamopoulos2017,Maskara2018,Chamberland2018,
  Fosel2018,Davaasuren2018,Baireuther2018,
  Breuckmann2018,Ni2018,Sweke2018}. But near optimal (or very fast
suboptimal) decoding algorithms are already proposed for these codes
\citep{BSV14a,Darmawan2018,Delfosse2017,Duclos-Cianci2010a}, which
exploit their regular lattice structure.  In contrast, for quantum
LDPC codes, which are defined on random graphs, only recently has a
decoding algorithm been found for the special family of expander codes
\citep{Leverrier2015,Fawzi2017,Grospellier} and the general case
remains open. Our main motivation to study this problem is that
quantum LDPC codes have the potential of greatly reducing the overhead
required to realize robust quantum processors \cite{G13b,Fawzi2018}.

In this paper, we train neural BP (NBP) decoders for quantum LDPC
codes. To guide the learning process, we construct a loss function
that takes into account error degeneracy. We present results for the
toric code \citep{Kitaev2006a}, the quantum bicycle code
\citep{MacKay2004} and the quantum hypergraph-product code
\citep{Gilles2009}. Decoding accuracy improves up to 3 orders of
magnitude compared with the untrained BP decoder, and the improvement
is even more substantial when we ignore detected but uncorrected
errors. While we do not completely solve the LDPC decoding problem here, our results suggest that an important step forward was realized, and the strategy could be applied more broadly, for instance in many-body physics. That general strategy consists in training a neural network to solve a quantum problem, with initial conditions corresponding to the BP algorithm that solves the classical counterpart.

\emph{LDPC codes.---} A linear error-correcting code can be
represented by its parity-check matrix $H$ with binary (0 or 1) matrix
elements.  Codewords $\mathbf{c}$'s satisfying $H\mathbf{c}=\mathbf{0}
\mod2$. As a result, when an error pattern $\mathbf{e}$ is imposed on
the codeword $\mathbf{c}\to\mathbf{c}'=\mathbf{c}+\mathbf{e}\mod2$,
there will be a measurable syndrome pattern
$\mathbf{s}=H\mathbf{c}'=H\mathbf{e}\mod2$, which signals the
occurrence of the error $\mathbf{e}$. The role of the decoder is to
infer the error pattern $\mathbf{e}$ from the measured syndrome
pattern $\mathbf{s}$. Classical LDPC codes are error-correcting codes
with sparse parity-check matrices, i.e., where the number of 1' in
each column and row are bounded by constants independent of the matrix
size.

\emph{Belief propagation.---} The Tanner graph is a graphical
representation of the parity-check matrix $H$, with a set of variable
nodes $\left\{ e_{v}|v=1,\ldots,n\right\} $ (containing the error
pattern) and a set of check nodes $\left\{ s_{c}|c=1,\ldots,m\right\}
$ (containing the syndrome pattern). There is an edge between $e_{v}$
and $s_{c}$ if $H_{cv}=1$. Neighborhoods of variables and checks are
defined by $\mathcal{N}\left(v\right)=\left\{ c|H_{cv}=1\right\} $ and
$\mathcal{N}\left(c\right)=\left\{ v|H_{cv=1}\right\} $, respectively.

Belief propagation (BP) is an iterative algorithm for approximating
the average value of each variable node $e_{v}$, over all error
patterns $\mathbf{e}$'s that are consistent with the given syndrome
pattern $\mathbf{s}$ (meaning $H\mathbf{e}=\mathbf{s}$). In performing
the average, each error pattern $\mathbf{e}$ is weighted by a
probability $P\left(\mathbf{e}\right)=\prod_{v}P\left(e_{v}\right)$,
which should accurately model the noise statistics of the physical
device carrying the information.  Mathematically speaking, BP solves
the posterior marginal probability for each variable node
$P\left(e_{v}=1|\mathbf{s}\right)\propto\sum_{\mathbf{e}/e_{v}}P\left(\mathbf{s}|\mathbf{e}/e_{v},e_{v}=1\right)P\left(\mathbf{e}/e_{v},e_{v}=1\right)$.
This goal is achieved by iterating the following simple BP equations:
\begin{gather}
  \mu_{v\to
    c}^{\left(t+1\right)}=l_{v}+\sum_{c'\in\mathcal{N}\left(v\right)
    \backslash c} \mu_{c'\to
    v}^{\left(t\right)},\label{eq:cvvc}\\ \mu_{c\to
    v}^{\left(t+1\right)}=\left(-1\right)^{s_{c}}2\tanh^{-1}
  \prod_{v'\in\mathcal{N}\left(c\right) \backslash
    v}\tanh\frac{\mu_{v'\to c}^{\left(t\right)}}{2},\label{eq:vccv}
\end{gather}
where
$l_{v}=\log\left(\frac{P\left(e_{v}=0\right)}{P\left(e_{v}=1\right)}\right)$
is the prior log-likelihood ratio for variable $e_{v}$ and
$\mathcal{N}\left(x\right) \backslash y$ is the set of all neighbors
of $x$ except for $y$ \citep{Richardson2008}.  The initial condition
for the iteration is $\mu_{v\to c}^{\left(t=0\right)}=0$, and after
$T$ steps (sufficiently long), one stops the iteration and performs
the following marginalization for the posterior log-likelihood ratio:
\begin{equation}
  \mu_{v}=l_{v}+\sum_{c\in\mathcal{N}\left(v\right)} \mu_{c\to
    v}^{\left(T\right)}.\label{eq:final}
\end{equation}
The posterior marginal probability relates to $\mu_{v}$ according to
$\log\left(\frac{P\left(e_{v}=0|\mathbf{s}\right)}{P\left(e_{v}=1|\mathbf{s}\right)}\right)=\mu_{v}$.
Equivalently
$P\left(e_{v}=1|\mathbf{s}\right)=\sigma\left(\mu_{v}\right)$ and
$P\left(e_{v}=0|\mathbf{s}\right)=1-\sigma\left(\mu_{v}\right)$, where
$\sigma\left(x\right)=1/\left(e^{x}+1\right)$ is the Fermi function
(or horizontally-flipped sigmoid function). The inferred error pattern
maximizes these marginal probabilities, i.e., $e_{v}$ is inferred to
be 0/1 when $\mu_{v}$ is positive/negative.

\begin{figure}
\includegraphics[width=85mm]{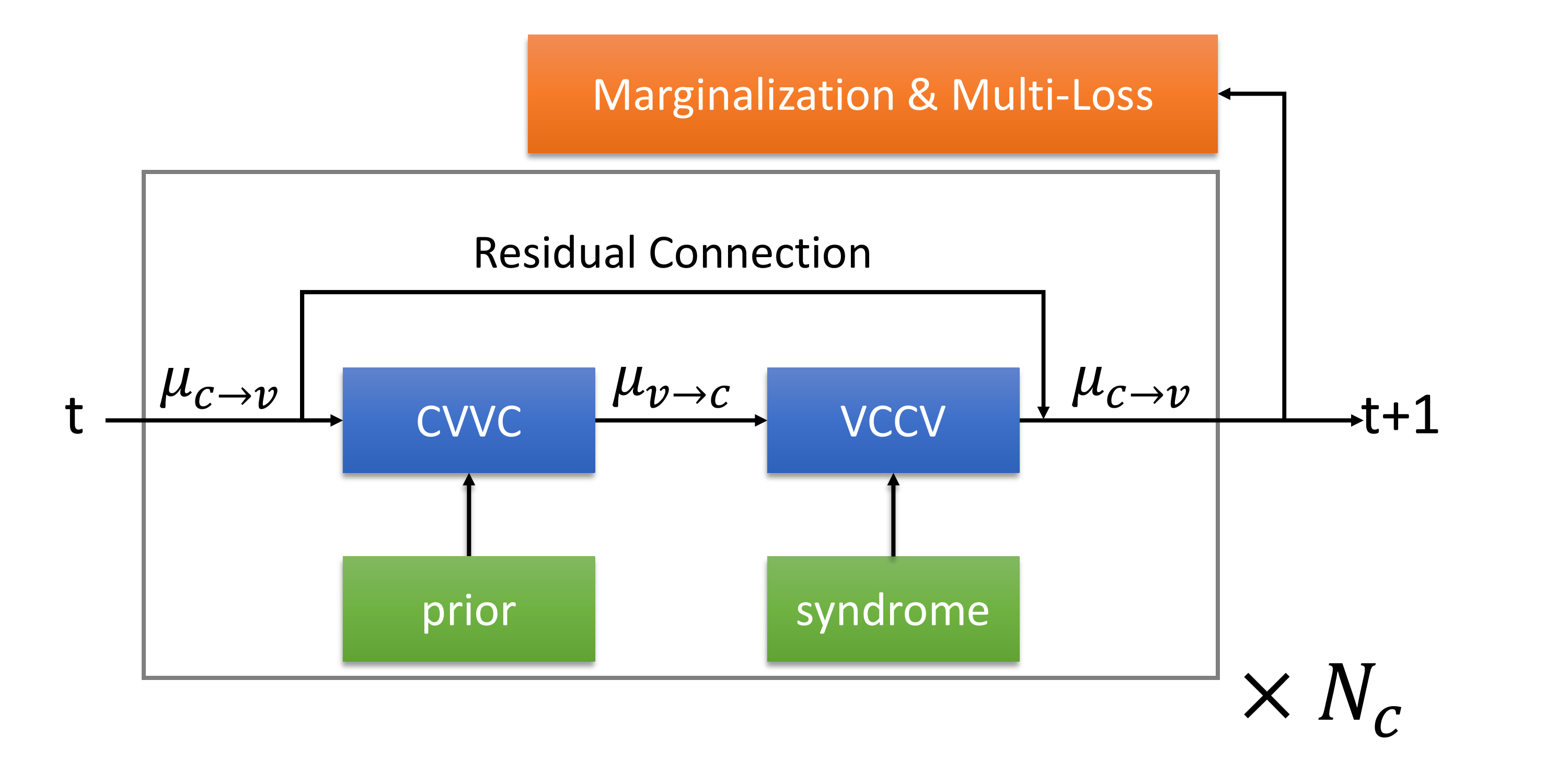}
\caption{ Schematics of the NBP decoder. The main cycle in the gray
  box is repeated $N_{c}$ times. Inside one cycle there are two phases
  of computation, the $cv\rightarrow vc$ and $vc\rightarrow cv$ phases
  which are governed by Eqs. (\ref{eq:neural-cvvc}) and
  (\ref{eq:neural-vccv}) respectively. The inputs to the neural
  network are denoted by the green boxes, where the prior and syndrome
  correspond to $\left\{ l_{v}\right\} $ in Eq. (\ref{eq:neural-cvvc})
  and $\left\{ s_{c}\right\} $ in Eq.  (\ref{eq:neural-vccv})
  respectively. After each cycle, the set of output  $\left\{
  \mu_{c\to v}\right\} $ is marginalized by
  Eq. (\ref{eq:neural-final}) and the resulting $\left\{
  \mu_{v}\right\} $ is sent to the loss function
  Eq. (\ref{eq:loss-quantum}). We also introduce residual connections
  to facilitate training of deep networks \citep{He2015b}. (See SM for
  details.)  \label{fig:schematics}}
\end{figure}

\emph{Neural belief propagation.---} The above iterative procedure can
be exactly mapped to a deep neural network, where each neuron
represents a message $\mu_{c\to v}$ or $\mu_{v\to c}$
\citep{Nachmani2016}.  (See Fig. \ref{fig:schematics}.) This permits
generalization of the original BP algorithm by introducing additional
``trainable'' weights $w_{c'v,vc}^{\left(t\right)}$ and
$w_{cv,v}^{\left(T\right)}$, and ``trainable'' biases
$b_{v}^{\left(t\right)}$ and $b_{v}^{\left(T\right)}$.  Specifically,
in this NBP algorithm, Eqs. (\ref{eq:cvvc}, \ref{eq:vccv},
\ref{eq:final}) are modified to:
\begin{gather}
\mu_{v\to
  c}^{\left(t+1\right)}=l_{v}b_{v}^{\left(t\right)}+\sum_{c'\in\mathcal{N}\left(v\right) \backslash c}\mu_{c'\to
  v}^{\left(t\right)}w_{c'v,vc}^{\left(t\right)},\label{eq:neural-cvvc}\\ a\left(\mu_{c\to
  v}^{\left(t+1\right)}\right)=i\pi
s_{c}+\sum_{v'\in\mathcal{N}\left(c\right) \backslash v}a\left(\mu_{v'\to
  c}^{\left(t\right)}\right),\label{eq:neural-vccv}\\ \mu_{v}=l_{v}b_{v}^{\left(T\right)}+\sum_{c\in\mathcal{N}\left(v\right)}\mu_{c\to
  v}^{\left(T\right)}w_{cv,v}^{\left(T\right)},\label{eq:neural-final}
\end{gather}
respectively \citep{Wymeersch2011,Nachmani2016}. Notice that all
equations above have the form of weighted sum plus bias, interleaved
with the nonlinear function
$a\left(x\right)=\log\left(\tanh\left(x/2\right)\right)$.  This is the
canonical form of feed-forward neural networks
\citep{goodfellow2016deep}.  When setting all newly introduced
parameters to 1, these equations became the standard BP equations \footnote{For numerical stability, argument of $\tanh^{-1}(\cdot)$  is truncated to $(-1+\epsilon,1-\epsilon)$ and we choose $\epsilon=10^{-4}$. Decreasing $\epsilon$ does not change the main conclusion of the paper.}.

To train these weights, one minimizes a carefully designed loss
function $\mathcal{L}$ by back-propagating its gradients w.r.t. all
trainable parameters. E.g., biases are updated according to $\Delta
b_v^{(t)}=-lr\times\partial{\mathcal{L}}/\partial{ b_v^{(t)}}$, where
$lr$ is the learning rate. For classical codes, one aims for
reproducing the whole error pattern exactly, so the natural choice of
the loss function is the binary cross entropy function between the inferred error pattern and the true error pattern:
\begin{equation}
\mathcal{L}\left(\vec{\mu};\mathbf{e}\right)=-\sum_{v}e_{v}\log\sigma\left(\mu_{v}\right)+\left(1-e_{v}\right)\log\left[1-\sigma\left(\mu_{v}\right)\right].\label{eq:loss-classical}
\end{equation}

\emph{Quantum setting.---}
Quantum noise can be modeled by 
random Pauli operators $I$, $\hat X$, $\hat Y$, and $\hat Z$ on the qubits. A convenient way of bookkeeping a $N$-qubit error
 uses a $2N$-bit string $\mathbf{e}$ representing the 
Pauli operator: $\hat{\mathcal{P}}\left(\mathbf{e}\right)=\prod_{1\le i\le N}\left[\hat{X}_{i}\right]^{e_{i}}\left[\hat{Z}_{i}\right]^{e_{i+N}}$. In this representation,
two Pauli-strings operators $\hat{\mathcal{P}}\left(\mathbf{a}\right)$
and \textbf{$\hat{\mathcal{P}}\left(\mathbf{b}\right)$} commute/anticommute
when $\mathbf{a}^{T}M\mathbf{b}$ is even/odd, where the symplectic inner product is defined with  $M=\left(\begin{array}{cc}
 & 1_{N\times N}\\
1_{N\times N}
\end{array}\right)$. Note that all Pauli-string operators satisfy $\hat{\mathcal{P}}^{2}=1$.

Likewise, the quantum codewords $\ket\psi$ are defined by a set of constraints $S_j\ket\psi = +\ket\psi$ where each stabilizer generator $S_j$ is a Pauli-string operator. For these equations to have a solution, it is necessary for the $S_j$ to mutually commute and to not generate $-1$ under multiplication. Using the above bookkeeping, we can represent each stabilizer generator $S_j$ by a $2N$-bit string, and assemble these strings as rows of a parity-check matrix $H$. A quantum LDPC code is one whose parity-check matrix is row- and column-sparse.

There is a crucial difference between classical and quantum error
correction. In the classical case, successful decoding means the inferred
error $\mathbf{e}^{\mathrm{inf.}}$ is exactly the same as the true error $\mathbf{e}$; while in the quantum
case, one only requires the total error $\mathbf{e}^{\mathrm{tot.}}=\mathbf{e}+\mathbf{e}^{\mathrm{inf.}}\mod2$
to belong to the ``stabilizer group'' -- the set of all Pauli-string
operators spanned by the rows of \textbf{$H$}. This is because two Pauli-string operators $E$ and $F=ES_j$ that differ by a stabilizer have identical action on all codestates. To test if $\mathbf{e}^{\mathrm{tot.}}$
belongs to the stabilizer group, one simply needs to check that it commutes with all the operators that commute with the stabilizers, i.e., that $H^\perp M \mathbf{e}^{\mathrm{tot.}}=\mathbf{0}\mod2$ where $H^\perp$ is the matrix that generates the orthogonal complement of $H$ with respect to the symplectic inner product, $HM\left(H^\perp\right)^T = {\bf 0}\mod2$.


The above analysis motivates the design the following loss function
tailored for quantum error correction:
\begin{gather}
\mathcal{L}\left(\vec{\mu};\mathbf{e}\right)=\sum_{i}f\left(\sum_{jk}H^\perp_{ij}M_{jk}\left[e_{k}+\sigma\left(\mu_{k}\right)\right]\right).\label{eq:loss-quantum}
\end{gather}
Note the parity check $parity\left(x\right)=x\mod2$ is replaced by
the continuous and differentiable function $f\left(x\right)=\left|\sin\left(\pi x/2\right)\right|$
to facilitate gradient-based machine-learning techniques. This loss
is minimized when the true error $\mathbf{e}$ and the inferred error
$\mathbf{e}^{\mathrm{inf.}}$ sum to a stabilizer. 

The loss function can also be averaged over all NBP-cycles $\mathcal{\bar{L}}=\frac{1}{N_{c}}\sum_{i=1}^{N_{c}}\mathcal{L}\left(\vec{\mu}^{\left(i\right)};\mathbf{e}\right),$
which requires marginalization after each cycle. In this work we use
a variation of this form. See SM for more details.

\begin{figure}
\includegraphics[width=85mm]{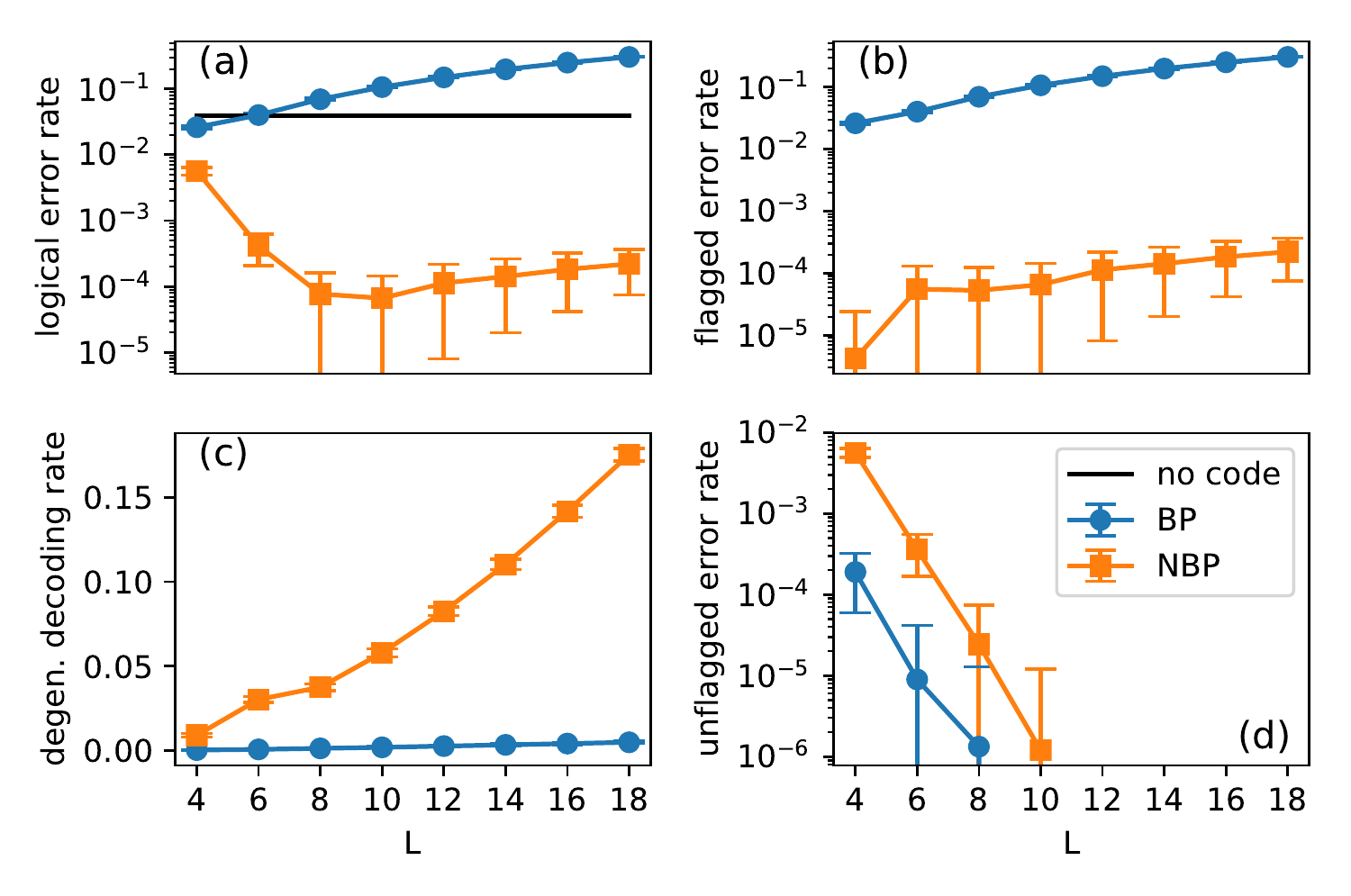}

\caption{Training the NBP decoder for the toric code with different code sizes.
(a) The logical error rate decreases substantially after training
(tested at $p_{err}=0.01$). Here the logical error rate is broken
up into two terms in (b) and (d), corresponding to ``flagged'' and
``unflagged'' errors, respectively. (See main text for details.)
(c) The NBP decoder exploits degeneracy by correctly decoding
with an error pattern that is not exactly the same as the true error
pattern. Training parameters: $N_{c}=25$, $lr=2\times10^{-4}$. \label{fig:toric_code_1}}

\end{figure}
\begin{figure}
  \includegraphics[width=85mm]{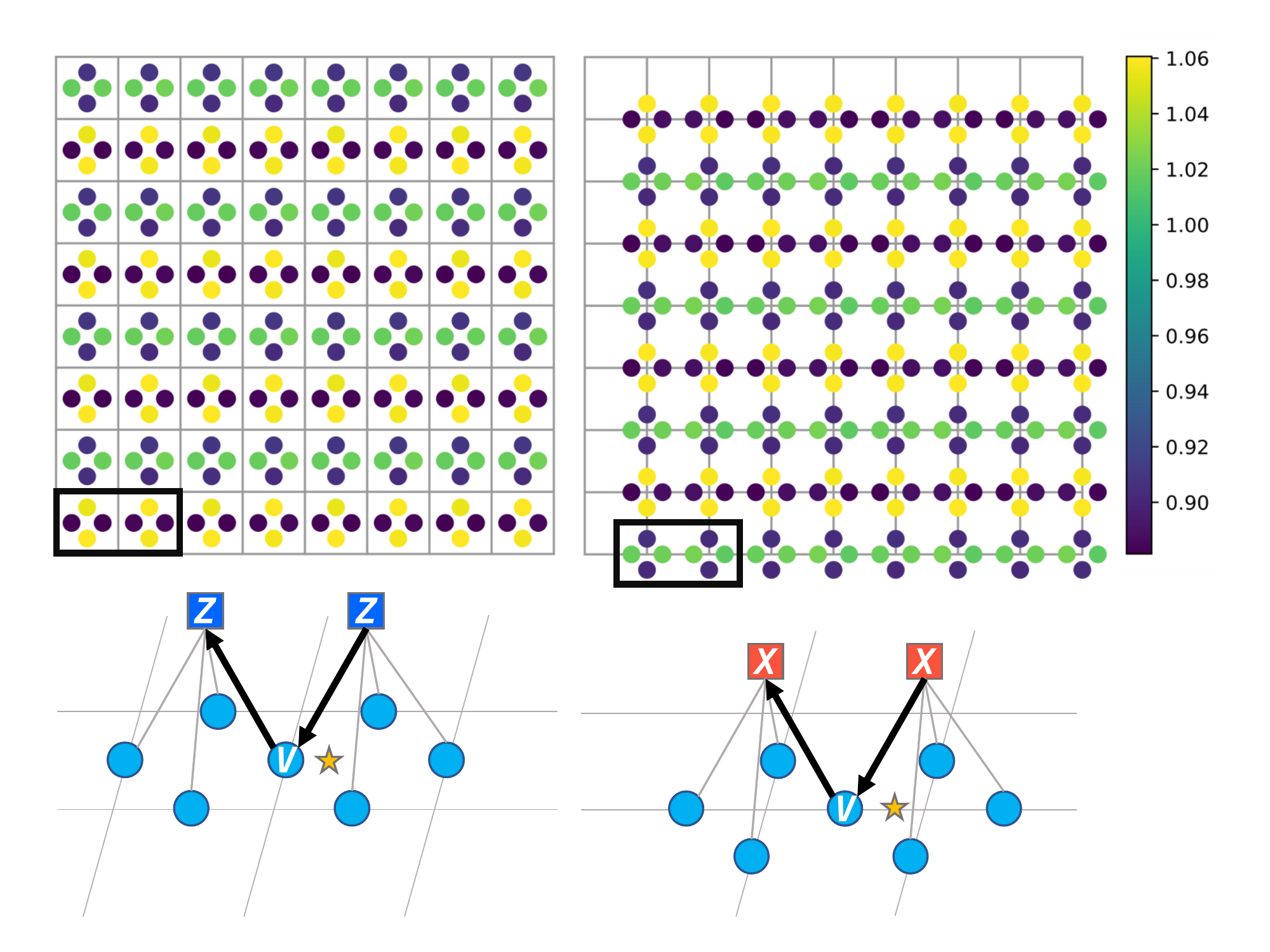}
  \caption{(Top panels) One of the observed patterns of the learned
    weights for the toric code. To obtain this clear pattern, we
    implemented weights sharing (see main text) with
    $\mathbf{G}=\left(2i,2j\right),i,j=0,1,\ldots$ A noisy version of
    the same pattern is observed when weight sharing is turned
    off. (Bottom panels) Tanner graph of the toric code and the
    position of $cvvc$ weights (yellow star). Correspondence to the
    top panels is marked by the black box. \label{fig:cvvc} }
\end{figure}

\textit{Toric code.---} We first study the toric code
\citep{Kitaev2006a} on an $L\times L$ square lattice, which is a
simple and widely studied quantum LDPC. (See Fig. \ref{fig:cvvc} for
the local Tanner graph.) During training, we generate error patterns
consisting of independent $X$ and $Z$ errors with physical error rate
$p_{err}$, i.e., $P\left(e_{v}=1\right)=p_{err}$ for all $v$. In each
minibatch, 120 error patterns are drawn from 6 physical error rates
that are uniformly distributed in the range
$p_{err}\in\left[0.01,0.05\right]$.  After $\sim10000$ minibatches, we
test the performance of the trained decoder. Figure
\ref{fig:toric_code_1} compares the original BP decoder (before
training) and the trained NBP decoder at $p_{err}=0.01$ for various
code sizes. Training significantly enhances decoding accuracy up to
three orders of magnitude (Fig.~\ref{fig:toric_code_1}a), and we
observe that the training time required for convergence depends weakly
on the code size $L$. (See SM for details.)

We can distinguish two types of decoding failure. ``Flagged'' failures
occur when the correction inferred by the decoder does not return the
system to the code space -- there remains a non-trivial syndrome after
decoding.  ``Unflagged'' failures occur when the correction return the
system to the wrong code state. These two contributions to the overall
logical error rate are shown in Fig. \ref{fig:toric_code_1}b and
\ref{fig:toric_code_1}d, respectively. We observe that training
greatly reduces flagged failures at the expense of slightly increasing
unflagged failures, and overall there is a significant net decrease of
failures.  It should be noted that flagged failures are benign because
they can be re-decoded, using either a more accurate but more
expensive decoder (e.g. the minimum-weight perfect matching
\cite{E65a}) or a higher layer of code for erasure errors.  Such a
mixed decoding strategy would combine the speed and flexibility of BP
decoder and reliability of a more expensive decoder used on a very
small fraction (e.g. $10^{-4}$) of instance.

The loss function Eq. (\ref{eq:loss-quantum}) takes into account error
degeneracy, and we see on Fig.  \ref{fig:toric_code_1}c that the
frequency of successful decoding where the actual and the inferred
error differ by a stabilizer increases with the code length.  This
rate was nearly zero with the untrained decoder (see SM for examples
of learned stabilizers).

The periodic nature of the toric code inspired us to utilize a
weight-sharing technique, where the weights
$w_{c_{1}'v_{1},v_{1}c_{1}}$ and $w_{c_{2}'v_{2},v_{2}c_{2}}$ are
invariant under lattice translation $\mathbf{G}$. We can control the
amount of sharing by the size of $\mathbf{G}$ (similar to the
filter-size in convolutional neural networks). Fig. \ref{fig:cvvc} is
a graphical representation of the trained weights, and suggests that
symmetry breaking improves BP for quantum codes. We also observe that
weights trained on one code size can also increase the performance
when applied to codes of different sizes, which implies that the
learning is universal/transferable (see SM for more details).

\begin{figure}
\includegraphics[width=85mm]{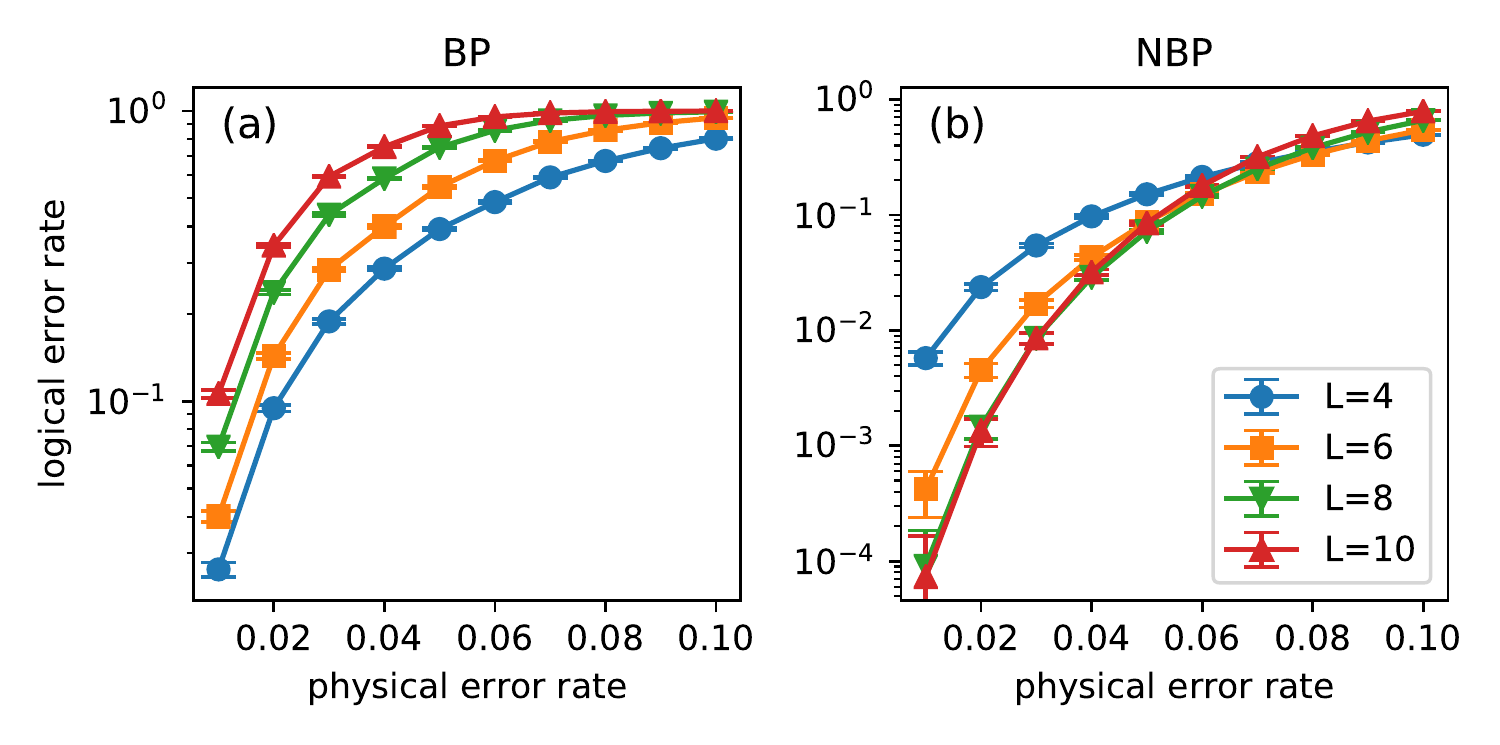}

\caption{Evolution of the logical error rate as a function of the
  physical error rate, for the BP/NBP decoder before/after
  training. (a) Before training the performance of the BP decoder is
  worse for larger code sizes at all physical error rates. (b) After
  training, the performance improves substantially for all code sizes
  at all physical error rates, and the performance curves start to
  cross each other. This indicates the development of a
  threshold. \label{fig:toric_code_2}}
\end{figure}

Figure \ref{fig:toric_code_2} shows that significant improvement can
be achieved across a range of physical error rates. Using the original
BP, increasing the code size leads to worse performance. After
training, performance improves with size for sufficiently low error
rates, and the trend indicates that further improved training might
lead to a BP decoder with a finite threshold.

When the neural network is initialized away from BP, training gets
stuck at a much worse local minimum. This illustrates the importance
of incorporating domain knowledge (when possible) before using general
machine-learning methods as black boxes, which contrasts with prior
uses of neural net decoding of the toric code
\cite{Torlai2017a,Krastanov2017}.

\begin{figure}
\includegraphics[width=85mm]{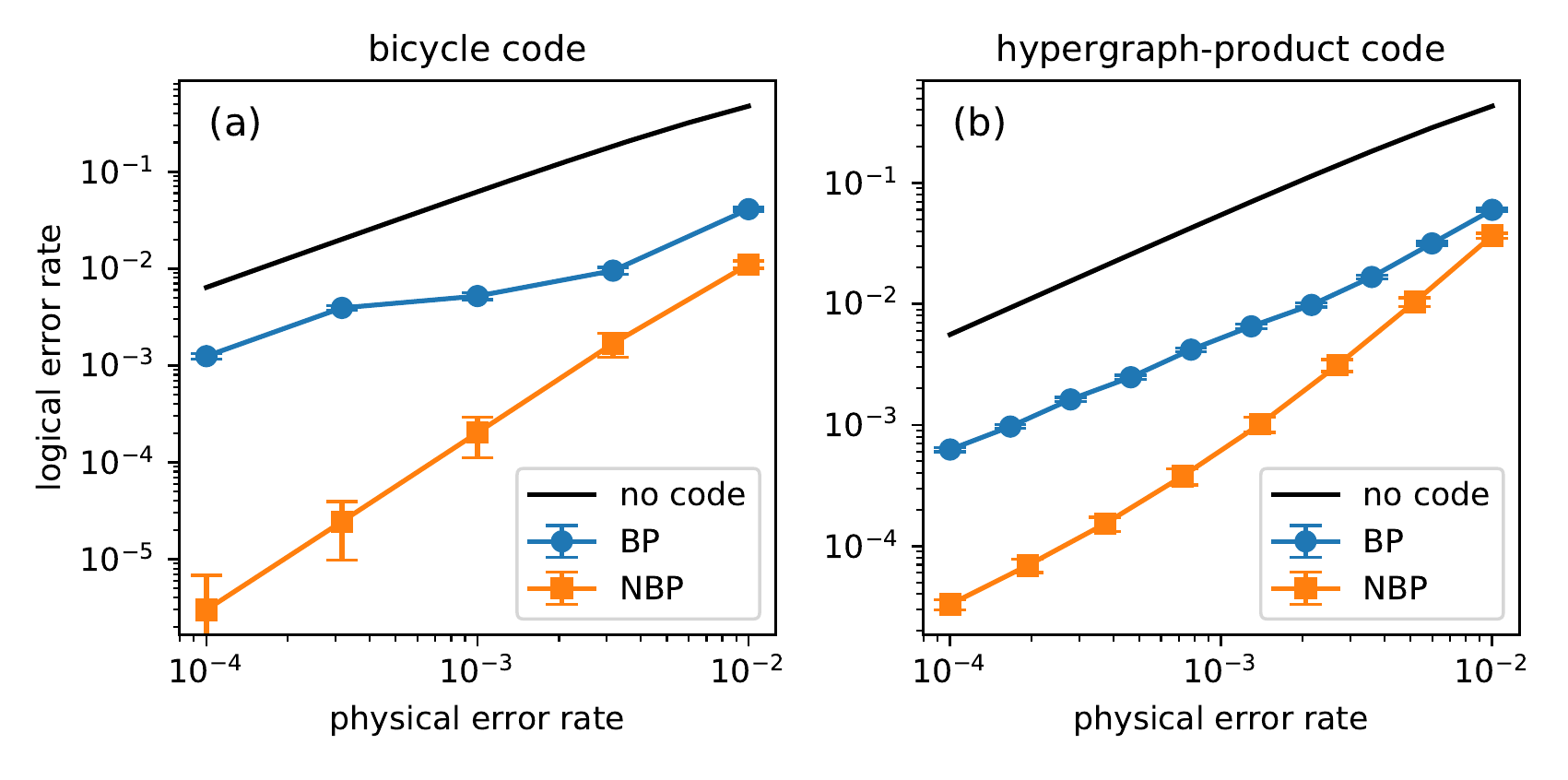}\caption{Training greatly improves
  the BP decoder for quantum LDPC codes with high rates. (a) Quantum
  bicycle code with code parameter $\left[\left[256,32\right]\right]$
  and rate $r=0.125$. Training parameters: $N_{c}=12$,
  $lr=1\times10^{-4}$.  (b) Quantum hypergraph-product code with code
  parameter $\left[\left[129,28\right]\right]$ and rate
  $r\sim0.2$. Training parameters: $N_{c}=12$, $lr=1\times10^{-4}$.
\label{fig:LDPC}}
\end{figure}
\textit{Quantum LDPC codes with high rate.---} The toric code encodes
a constant number $K=2$ of qubits in a growing number $N$ of physical
qubits, thus achieving a vanishing rate $r=K/N$. We now turn to
quantum LDPC codes with constant rates.

The quantum bicycle code \citep{MacKay2004} is a quantum LDPC code
constructed from a random binary vector $\mathbf{A}$ of size $N/2$.
First, all cyclic permutations of $\mathbf{A}$ are collected as
columns in a matrix $C$. Then $C$ is concatenated with its transpose
to form $H_{0}=\left[C,C^{T}\right]$, from which $K/2$ rows are chosen
randomly and removed. After these constructions, $H_{0}$ is a
self-dual matrix (meaning $H_{0}H_{0}^{T}=0\mod2$) of size
$\left(N-K\right)/2\times N$.  The final parity-check matrix for the
quantum bicycle code is $H=\left(\begin{array}{cc} H_{0}\\ & H_{0}
\end{array}\right)$.
The sparsity of this matrix can be controlled by the number of nonzero
elements in $\mathbf{A}$. Training the NBP decoder for a quantum
bicycle code with $N=256$, $K=32$ and $\sum_{i}A_{i}=8$ improves the
accuracy up to 3 orders of magnitude (Fig. \ref{fig:LDPC}a).

The quantum hypergraph-product code \citep{Gilles2009} is constructed
from two classical codes with parity-check matrices
$\left[H_{1}\right]_{m_{1}\times n_{1}}$ and
$\left[H_{2}\right]_{m_{2}\times n_{2}}$. The following products are
constructed $H_{X}=\left[\begin{array}{cc} H_{1}\otimes I_{n_{2}\times
      m_{2}}, & I_{m_{1}\times n_{1}}\otimes
    H_{2}^{T}\end{array}\right]$ and $H_{Z}=\left[\begin{array}{cc}
    I_{n_{1}\times m_{1}}\otimes H_{2} & ,H_{1}^{T}\otimes
    I_{m_{2}\times n_{2}}\end{array}\right]$, and the parity-check
matrix of the quantum code follows $H=\left(\begin{array}{cc}
  H_{X}\\ & H_{Z}
\end{array}\right)$,
which performs $m=m_{1}n_{2}+n_{1}m_{2}$ checks on
$n=m_{1}m_{2}+n_{1}n_{2}$ qubits. In this paper, we study a
hypergraph-product code, for which $H_{1}$ and $H_{2}$ are the
classical $\left[7,4,3\right]$ and $\left[15,7,5\right]$ BCH codes,
respectively. This code has rate $r=28/129\sim0.2$. Training the NBP
decoder for this code improves the accuracy up to one order of
magnitude (Fig. \ref{fig:LDPC}b).

\emph{Conclusions.---} We significantly improved the
belief-propagation decoders for quantum LDPC codes by training them as
deep neural networks.  Our results on the toric code, the quantum
bicycle code and the quantum hypergraph-product code all show orders
of magnitude of enhancement in decoding accuracy. The original belief
propagation is known to have bad performance for quantum
error-correcting codes \cite{Poulin2008}. On the other hand, training
a neural decoder with general architecture has been reported to be
hard for large codes \cite{Gruber2017,Maskara2018}.  Our results
indicate that combining the general framework of machine learning and
the specific domain knowledge of quantum error correction is a
promising approach, when neither works well individually.

The significance of this result is supported by the tremendous success
of BP with classical LDPC codes \citep{Richardson2008}, and the fact
that quantum LDPC codes promise a low-overhead fault-tolerant quantum
computation architecture \cite{G13b}. In addition, our techniques
could be adapted to uses of BP in other quantum many-body problems,
such as improving the quantum cavity method
\citep{Hastings2007,Poulin2008a,Laumann2008,Poulin2011b}.

\begin{acknowledgments}
\emph{Acknowledgments.---} Y.H.L would like to thank helpful
discussions with Pavithran Iyer, Anirudh Krishna, Xin Li, Alex Rigby, and Colin
Trout. Special thanks go to Liang Jiang and Stefan Krastanov, who
shared similar ideas. This research was undertaken thanks in part to
funding from the Canada First Research Excellence Fund. Computations
were made on the supercomputer Helios managed by Calcul Qu\'ebec and
Compute Canada. The operation of this supercomputer is funded by the
Canada Foundation for Innovation (CFI), the minist\`ere de
l'\'Economie, de la science et de l'innovation du Qu\'ebec (MESI) and
the Fonds de recherche du Qu\'ebec - Nature et technologies (FRQ-NT). 
We used {\tt TensorFlow} \cite{Abadi2016} to build and train neural belief-propagation decoders.
\end{acknowledgments}

\end{document}